\documentclass[journal]{IEEEtran}
\ifCLASSINFOpdf
  \usepackage[pdftex]{graphicx}
\else
\fi
\usepackage{amsmath}
\usepackage{algpseudocode}
\usepackage{algorithm}
\begin{document}
%
\title{Pseudoinverse Diffusion Models for Generative CT Image Reconstruction from Low Dose Data}
%
%
%

\author{Matthew~Tivnan,
Dufan~Wu, and~Quanzheng~Li,
\thanks{All authors were with the Department
of Radiology, Massachusetts General Hospital and Harvard Medical School and , Boston,
MA, 02114 USA.}
}

\maketitle


\begin{abstract}

Score-based diffusion models have significantly advanced generative deep learning for image processing. Measurement conditioned models have also been applied to inverse problems such as CT reconstruction. However, the conventional approach, culminating in white noise, often requires a high number of reverse process update steps and score function evaluations. To address this limitations, we propose an alternative forward process in score-based diffusion models that aligns with the noise characteristics of low-dose CT reconstructions, rather than converging to white noise. This method significantly reduces the number of required score function evaluations, enhancing efficiency and maintaining familiar noise textures for radiologists, Our approach not only accelerates the generative process but also retains CT noise correlations, a key aspect often criticized by clinicians for deep learning reconstructions.  In this work, we rigorously define a matrix-controlled stochastic process for this purpose and validate it through computational experiments. Using a dataset from The Cancer Genome Atlas Liver Hepatocellular Carcinoma (TCGA-LIHC), we simulate low-dose CT measurements and train our model, comparing it with a baseline scalar diffusion process and conditional diffusion model. Our results demonstrate the superiority of our pseudoinverse diffusion model in terms of efficiency and the ability to produce high-quality reconstructions that are familiar in texture to medical professionals in a low number of score function evaluations. This advancement paves the way for more efficient and clinically practical diffusion models in medical imaging, particularly beneficial in scenarios demanding rapid reconstructions or lower radiation exposure.
\end{abstract}


%
\IEEEpeerreviewmaketitle


\section{Introduction}

Score-based diffusion models have been introduced as a powerful framework for deep generative modeling and have revolutionized the field of image generation \cite{song2020score}. These models are based on a forward stochastic process where an image is progressively degraded in quality. Then, a score-matching neural network can be used to approximate the reverse time stochastic process, generating high-quality images from degraded inputs. This approach has been particularly successful due to its ability to model complex data distributions and generate images with high fidelity. We propose a novel forward process for score-based diffusion models that converges not to white noise, but to low-dose CT reconstructed images. This approach offers a more direct and efficient path to high-quality image generation. Our method reduces the computational burden, requiring fewer evaluations of the score function. Additionally, the intermediate images correspond to reconstructions from intermediate radiation dose levels in terms of the noise correlations and magnitude.

There have been many recent studies investigating the application of diffusion models to solving noisy inverse problems. Conditional diffusion models leverage an auxiliary input for measurements to facilitate the generation of high-quality images from lower-quality measurements \cite{batzolis2021conditional}. However, these models typically conclude their diffusion process in white noise, contrasting our method which effectively ends with a pseudoinverse low dose reconstruction. Another method for measurement conditioned diffusion for solving inverse problems is Diffusion Posterior Sampling (DPS) \cite{chung2022diffusion}. DPS has been successfully applied to CT reconstruction in various studies \cite{xia2023diffusion, li2023diffusion, huang2023proceedings}, presenting a framework that combines a pre-trained prior model with a physics-driven measurement likelihood. Unlike these approaches, our model is trained using paired data and supervised learning. Instead of modifying the unconditional reverse process with a measurement likelihood term, our method commences the reverse process  directly with reconstructed images to produce posterior samples. Come-Closer-Diffuse-Faster (CCDF) approach addresses a critical downside of traditional diffusion models — their inherently slow sampling rate \cite{chung2022come}. CCDF circumvents this by initiating the reverse process from a point that is not pure Gaussian noise, but a noisy version of the target image, thereby reducing the number of sampling steps required. While this approach shows improvements in efficiency, it relies on an initialization which is the ground truth plus white noise, which is not immediately available. In our method, we use the same noise correlations inherent in CT reconstructions. This distinction underlines the efficiency and practical applicability of our method in the medical imaging domain.

Our work is most closely aligned with our previous work on Fourier Diffusion Models \cite{tivnan2023fourier}. These models replace scalar operations in the forward process with shift-invariant convolutions and additive stationary noise, enabling control over Modulation Transfer Function (MTF) and Noise Power Spectrum (NPS). However, Fourier Diffusion Models are restricted to shift invariant blur and stationary noise, which is not realistic for CT. In this work, we extend this concept to encompass realistic noise correlations found in CT reconstructions.


\section{Methods}

\subsection{Nonlinear CT Measurement Likelihood Model}

A statistical model of CT measurements, $\mathbf{Y}$, given true attenuation images, $\mathbf{X}$, is defined by the following conditional probability density function


\begin{gather}
    p(\mathbf{y}|\mathbf{x}) =  \mathcal{P}\Big(\mathbf{y}; \mathbf{\bar{y}}(\mathbf{x})  \Big)\label{p_y_given_x_poisson}\\
    \mathbf{\bar{y}}(\mathbf{x}) =  I_0 \exp{\Big(-\mathbf{A} \mathbf{x}\Big)} 
\end{gather}


\noindent where $\mathbf{x} \in \Re^{(N\times 1)}$ is the true image of linear attenuation coefficients, $\mathbf{A} \in \Re^{(M\times N)}$ is the forward projector, which we assume is a full-rank matrix with $M>N$, $I_0 \in \Re$ is the expected number of incident photons per pixel per view, $\mathbf{y} \in \Re^{(M \times 1)}$ is the vector of CT measurements assuming an ideal photon counting detector, and $\mathcal{P}(\mathbf{y}; \mathbf{\bar{y}}(\mathbf{x}))$ is the probability density function of a jointly independent Poisson random vector with expectation, $\mathbf{\bar{y}}(\mathbf{x}) \in \Re^{{(M \times 1)}}$ and covariance matrix, $D\{\mathbf{\bar{y}}(\mathbf{x})\}$. 




\begin{figure*}
    \centering
    \includegraphics[width=\textwidth]{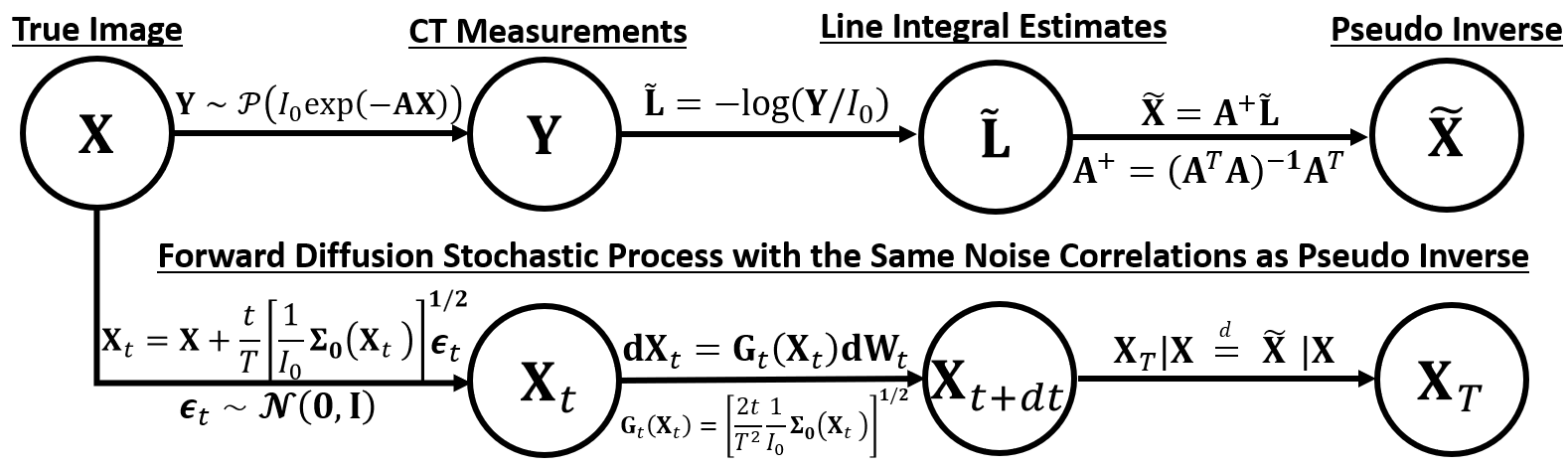}
    \caption{Probabilistic graphical model of pseudoinverse CT reconstruction and forward diffusion stochastic process. All probability density functions displayed in this figure are rigorously defined in the sections to  follow. The final time step of the forward diffusion process has the same distribution as pseudoinverse reconstructions, so the reverse diffusion process can be initialized with the pseudoinverse. }
    \label{fig:enter-label}
\end{figure*}
\subsection{Line Integral Estimation with Negative Logarithm}

It is a common preprocessing step to estimate attenuation line integrals by taking the negative logarithm of the gain corrected measurements, $\tilde{\mathbf{L}} = -\log(\mathbf{Y}/I_0)$. In our probabilistic framework, this deterministic data processing operation is represented by the following Dirac delta distribution,

\begin{equation}
    p(\boldsymbol{\tilde{\ell}}|\mathbf{y}) =  \delta\Big(\boldsymbol{\tilde{\ell}} - (-\log(\mathbf{y}/I_0)) \Big) .
\end{equation}

\noindent  We can analyze the mean signal and noise covariance transfer through the nonlinear elementwise logarithm operation using a first order Taylor approximation centered on $\mathbf{y}/I_0$. Using second order statistics, we can approximate the random vector of unitless line integral estimates with the following conditional Gaussian probability density function, 


\begin{equation}
    p(\boldsymbol{\tilde{\ell}}|\mathbf{x}) \approx   \mathcal{N} \Big(\boldsymbol{\tilde{\ell}}; \mathbf{A}\mathbf{x}, D\{1/I_0\exp{(-\mathbf{A}\mathbf{x})}\} \Big) . \label{p_l_tilde_given_x0}
\end{equation}

\noindent  This approximation is appropriate for many practical applications in CT imaging. Figure 1 shows that the noise induced bias for unitless attenuation line integral estimates is less than 0.01 if the expected photon counts are greater than 100. 

\begin{figure}
    \centering
    \includegraphics[trim={0 3mm 0 0},clip,width=0.49\textwidth]{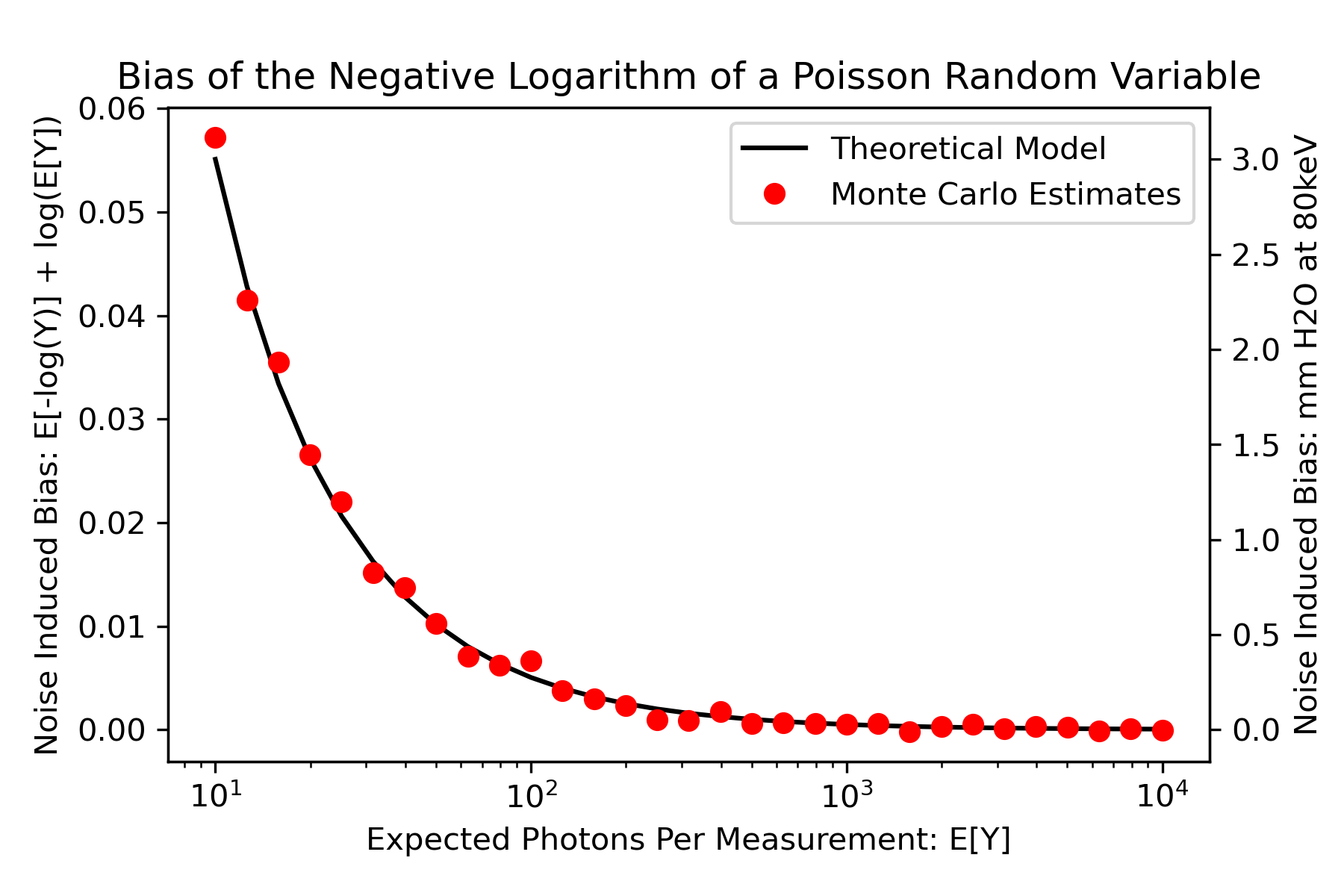}
    \caption{Expected error due to negative logarithm applied to a Poisson distributed random variable. Theoretical model of expected error was computed via numerical integration of the log Poisson distribution. Monte Carlo estimates were computed via sample expectation of 10,000 Poisson samples. }
    \label{fig:enter-label}
\end{figure}


\subsection{Pseudoinverse CT Image Reconstruction}

For the purposes of this work, we assume the CT reconstruction algorithm exactly implements the Moore-Penrose pseudoinverse, $\mathbf{\tilde{X}} = \mathbf{A}^{+} \mathbf{\tilde{L}} = (\mathbf{A}^T\mathbf{A})^{-1} \mathbf{A}^T \mathbf{\tilde{L}}$. Since $\mathbf{A}$ is defined as full-rank and $M>N$, we know that $\mathbf{A}^T\mathbf{A}$ is invertible. In the future, we are interested to extend our methods to include sparse CT acquisitions where $\mathbf{A}^T\mathbf{A}$ is not invertible, but we take this well conditioned case as a starting point. The pseudoinverse reconstruction is a deterministic function of line integrals, so it is modeled by the Dirac delta distribution,

\begin{gather}
    p(\mathbf{\tilde{x}}|\boldsymbol{\tilde{\ell}}) =  \delta\Big(\mathbf{\tilde{x}} - \mathbf{A}^{+}  \boldsymbol{\tilde{\ell}}\Big) =  \delta\Big(\mathbf{\tilde{x}} - (\mathbf{A}^T\mathbf{A})^{-1}\mathbf{A}^T \boldsymbol{\tilde{\ell}}\Big) .
\end{gather}

\noindent Further, the pseudoinverse is  a linear operator, and the input is a Gaussian distributed random vector defined in \eqref{p_l_tilde_given_x0}, so the output is also a Gaussian distributed random vector given by the conditional probability density function,

\begin{gather}
 p(\mathbf{\tilde{x}}|\mathbf{x}) \approx   \mathcal{N} \Big(\mathbf{\tilde{x}}; \mathbf{x},\frac{1}{I_0} \boldsymbol{\Sigma}_0(\mathbf{x}) \Big) \label{p_x_tilde_given_x0} \\
 \boldsymbol{\Sigma}_0(\mathbf{x}) \hspace{-1mm}= \hspace{-1mm}\mathbf{A}^{+} D\{1/\exp{(-\mathbf{A}\mathbf{x})}\} {\mathbf{A}^{+}}^T
\end{gather}

This end-to-end image quality model describes the correlated noise in pseudoinverse CT reconstructed images as a function of the dose, $I_0$, and the object-dependent spatial correlations as defined by, $\boldsymbol{\Sigma}_0(\mathbf{x})$. Note that the object dependent statistical weights and corresponding noise correlations can be well approximated by the raw measurements; that is, $\boldsymbol{\Sigma}_0(\mathbf{x}) \approx \boldsymbol{\Sigma}_0(\mathbf{\tilde{x}})=\boldsymbol{\Sigma}_0(\mathbf{A}^{+} \mathbf{y})$, which is a necessary property for our diffusion model derivations and similar to approximations made in several previous works \cite{erdogan2002monotonic} \cite{tilley2015model}.


\subsection{Forward and Reverse Diffusion Stochastic Process}

In this section, we define the continuous time stochastic process, $\mathbf{X}_t$ that begins with true images at $\mathbf{X}_0 = \mathbf{X}$ and ends with the same expectation and covariance as the pseudoinverse reconstructed images, $\mathbf{X}_T = \mathbf{\tilde{X}}$. To that end, we define the forward stochastic process as follows

\begin{equation}
\mathbf{X}_t | \mathbf{X}_0 = \mathbf{X}_0 + \frac{t}{T}[\frac{1}{I_0} \boldsymbol{\Sigma}_0(\mathbf{X}_0) ]^{1/2} \boldsymbol{\epsilon}_t, \quad \boldsymbol{\epsilon}_t = \frac{\mathbf{W}_t}{\sqrt{t}}  
\end{equation}

\noindent where $\boldsymbol{\Sigma}_0(\mathbf{X}_0)$ are the object dependent noise correlations defined in the previous section and $\mathbf{W}_t$ is the standard Wiener process with zero mean and covariance, $t \hspace{0.5mm}\mathbf{I}$, so $\boldsymbol{\epsilon}_t$ is zero mean and identity covariance white noise. Therefore, the noise standard deviation is proportional to $t$. The corresponding probability density function is

\begin{gather}
    p(\mathbf{x}_t|\mathbf{x_0}) = \mathcal{N}(\mathbf{x}_t; \mathbf{x}_0, \boldsymbol{\Sigma}_t) \label{p_x_t_given_x0} \\
    \boldsymbol{\Sigma}_t = \frac{t^2}{T^2} \frac{1}{I_0} \boldsymbol{\Sigma}_0(\mathbf{x}_0) 
\end{gather}

This forward diffusion stochastic process can be well approximated by the following stochastic differential equation, using Itô calculus notation,

\begin{gather}
\mathbf{dX}_t = \mathbf{G}_t(\mathbf{X}_t) \mathbf{dW}_t\\
\mathbf{G}_t(\mathbf{X}_t) = [\frac{2 t}{T^2 I_0} \boldsymbol{\Sigma_0}(\mathbf{X}_t) ]^{1/2} \approx  [\frac{2 t}{T^2 I_0} \boldsymbol{\Sigma_0}(\mathbf{X}_0) ]^{1/2} \label{G_t}
\end{gather}

\noindent where $\boldsymbol{\Sigma_0}(\mathbf{X}_t) \approx \boldsymbol{\Sigma_0}(\mathbf{X}_0)$ is a commonly used approximation for the object-dependent statistical weights and noise correlations as stated in the previous section. The corresponding conditional probability density function is 


\begin{equation}
    p(\mathbf{x}_{t+dt}| \mathbf{x}_{t}) = \mathcal{N}(\mathbf{x}_{t+dt}; \mathbf{x}_{t}, \frac{2 \hspace{0.5mm} t \hspace{0.5mm} dt }{T^2 I_0} \boldsymbol{\Sigma_0}(\mathbf{X}_t) ]^{1/2} )
\end{equation}


Following the general framework for score based generative models in \cite{song2020score}, based on the formula originally proposed by \cite{anderson1982reverse}, we can write the reverse diffusion stochastic process as


\begin{gather}
\mathbf{dX}_t = -\nabla \cdot \mathbf{G}_t(\mathbf{X}_t)\mathbf{G}_t(\mathbf{X}_t)^T dt  \\
 - \mathbf{G}_t(\mathbf{X}_t)\mathbf{G}_t(\mathbf{X}_t)^T \nabla \text{log} \hspace{0.5mm} p(\mathbf{X}_t) ]dt + \mathbf{G}_t(\mathbf{X}_t) \mathbf{d\bar{W}}_t  .\nonumber
\end{gather}

\noindent  We can ignore the divergence term, $\nabla \cdot \mathbf{G}_t(\mathbf{X}_t)\mathbf{G}_t(\mathbf{X}_t)^T$, because the derivative with respect to $\mathbf{X}_t$ is nearly zero, using the same assumption previously used in \eqref{G_t} that $\mathbf{G}_t(\mathbf{X}_t) \approx \mathbf{G}_t(\mathbf{X}_0)$. We also substitute a score-matching neural network, $\mathbf{s}_{\boldsymbol{\theta}}(\mathbf{x_t}, t)\approx  \nabla \text{log} \hspace{0.5mm} p(\mathbf{X}_t)$, for the score function 

\begin{equation}
\mathbf{dX}_t \hspace{-1.0mm} = \hspace{-1.0mm} -\mathbf{G}_t(\mathbf{X}_t)\mathbf{G}_t(\mathbf{X}_t)^T \mathbf{s}_{\boldsymbol{\theta}}(\mathbf{X_t}, t) dt + \mathbf{G}_t(\mathbf{X}_t) \mathbf{d\bar{W}}_t \label{reverse_sde}
 \end{equation}

This reverse process can be initialized directly with pseudoinverse reconstructed images, $\mathbf{\tilde{X}}$, and running the reverse process from $t=T$ to $t=0$ will result in a sample from the posterior distribution $\mathbf{X}|\mathbf{\tilde{X}}$. One important feature of our model is that the intermediate time points correspond to intermediate dose levels. This could be useful to control the tradeoff between noise and hallucinations through early stopping of the reverse process.

\begin{algorithm}
\caption{Pseudoinverse Diffusion Training}
\begin{algorithmic}
\State \textbf{Input:} $T$, $K$, $I_0$, $p(\mathbf{x})$
\State Initialize $\boldsymbol{\theta}$
\For{each epoch}
    \For{each sample}
        \State $\mathbf{X}_0 \sim p(\mathbf{x})$
        \State $t \sim \mathcal{U}(0,T)$ 
        \State $\boldsymbol{\tilde{\epsilon}}_t \sim \mathcal{N}(\mathbf{0},\mathbf{I})$ \# projection domain white noise
        \State $\mathbf{X}_t \leftarrow \mathbf{X}_0 + \frac{t}{T}\sqrt{\frac{1}{I_0}} \mathbf{A}^{+} D\{\frac{1}{\sqrt{\exp{(-\mathbf{A}\mathbf{X}_0)}}}\} \boldsymbol{\tilde{\epsilon}}_t$
        \State compute $\boldsymbol{\mu}_{\boldsymbol{\theta}}(\mathbf{X}_t, t)$
        \State optimizer step on training loss using \eqref{training_loss}
    \EndFor
\EndFor
\State \textbf{Return:} $\boldsymbol{\theta}$
\end{algorithmic}
\end{algorithm}
\begin{algorithm}
\caption{Pseudoinverse Diffusion Sampling (SDE)}
\begin{algorithmic}
\State \textbf{Input:} $T$, $\Delta t$, $I_0$, $\mathbf{Y}$
\State $\mathbf{\tilde{L}} \leftarrow -\text{log}(-\mathbf{Y}/{I_0})$
\State $\mathbf{\tilde{X}} \leftarrow \mathbf{A}^{+} \mathbf{\tilde{L}}$
\State $\mathbf{X}_T \leftarrow \mathbf{\tilde{X}}$
\For{each n in [N,0)}
    \State $\mathbf{G}_t(\mathbf{X}_t)\mathbf{G}_t(\mathbf{X}_t)^T \mathbf{s}_{\boldsymbol{\theta}}(\mathbf{X_t}, t) \leftarrow \frac{2}{t}(\mathbf{X}_t - \boldsymbol{\mu}_{\boldsymbol{\theta}}(\mathbf{X}_t, t))$
    \State $\mathbf{d\tilde{W}}_t \sim \mathcal{N}(\mathbf{0}, \Delta t \hspace{0.3mm} \mathbf{I} )$ \# projection domain white noise
    \State $\mathbf{G}_t(\mathbf{X}_t) \mathbf{d\bar{W}}_t \leftarrow \sqrt{\frac{2t}{T^2 I_0}} \mathbf{A}^{+} D\{\frac{1}{\sqrt{\exp{(-\mathbf{A}\mathbf{X}_t)}}}\} \mathbf{d\tilde{W}}_t$
    \State $\Delta \mathbf{X}_t \hspace{-1.0mm} \leftarrow \hspace{-1.0mm} -\mathbf{G}_t(\mathbf{X}_t)\mathbf{G}_t(\mathbf{X}_t)^T \mathbf{s}_{\boldsymbol{\theta}}(\mathbf{X_t}, t) \Delta t + \mathbf{G}_t(\mathbf{X}_t) \mathbf{d\bar{W}}_t $
    \State $\mathbf{X}_{t - \Delta t} \leftarrow \mathbf{X}_{t} - \Delta \mathbf{X}_t $
\EndFor
\State \textbf{Return:} $\mathbf{X}_0$
\end{algorithmic}
\end{algorithm}


\subsection{Training and Sampling Algorithms}

The score function for a Gaussian random vector is defined as $\nabla \text{log} \hspace{0.5mm} p(\mathbf{X}_t) = -\boldsymbol{\Sigma}_t^{-1} (\mathbf{X}_t - \text{E}[\mathbf{X}_t])$, so we use the following parameterization of the score matching neural network

\begin{equation}
\mathbf{s}_{\boldsymbol{\theta}}(\mathbf{X}_t, t) =  -\boldsymbol{\Sigma}_t^{-1} (\mathbf{X}_t - \boldsymbol{\mu}_{\boldsymbol{\theta}}(\mathbf{X}_t, t))
\end{equation}

\noindent Therefore, the first term of \eqref{reverse_sde} can be written as

\begin{equation}
-\mathbf{G}_t(\mathbf{X}_t)\mathbf{G}_t(\mathbf{X}_t)^T \mathbf{s}_{\boldsymbol{\theta}}(\mathbf{X_t}, t) = \frac{2}{t}(\mathbf{X}_t - \boldsymbol{\mu}_{\boldsymbol{\theta}}(\mathbf{X}_t, t))
\end{equation}

\noindent The advantage of this parameterization is that all  $\boldsymbol{\Sigma}_0(\mathbf{X}_t)$ and $I_0$ are cancelled out. The training loss function is 

\begin{equation}
E_{\mathbf{X_0}, \mathbf{X_t},   t} \Big[(\boldsymbol{\mu}_{\boldsymbol{\theta}}(\mathbf{X}_t, t)) - \mathbf{X}_0))^T \boldsymbol{\Sigma}_t^{-K} (\boldsymbol{\mu}_{\boldsymbol{\theta}}(\mathbf{X}_t, t)) - \mathbf{X}_0))\Big] \label{training_loss}
\end{equation}

For an L2 norm on mean estimates, one should set $K=0$ which results in uniform time dependent weights. For an L2 norm on score estimates, one should set $K=2$ which will result in time dependent weights proportional to $1/{t^4}$. Experimentally, we have found this places too small of a weight on times near $t=T$, and we choose the compromise $K=1$, which is log-likelihood loss with $1/{t^2}$ time dependence. The training algorithm is summarized in Algorithm 1. 

The second term of \eqref{reverse_sde}  requires a large matrix square root which would be computationally expensive. We propose an exact yet efficient formula to compute this stochastic term,

\begin{equation}
\mathbf{G}_t(\mathbf{X}_t) \mathbf{d\bar{W}}_t = \sqrt{\frac{2t}{T^2 I_0}} \mathbf{A}^{+} D\{\frac{1}{\sqrt{\exp{(-\mathbf{A}\mathbf{X}_t)}}}\} \mathbf{d\tilde{W}}_t
\end{equation}

\noindent where $ \mathbf{d\tilde{W}}_t \in \Re^{(M\times1)}$ is projection domain white noise with zero mean and covariance, $dt \mathbf{I}$. This represents applying elementwise statistical weights to the white noise in the projection domain followed by pseudoinverse reconstruction and scaling by time and dose. This formula is exact because the left and right sides of the equation are image domain Gaussian random vectors with zero mean and covariance, $dt \mathbf{G}_t(\mathbf{X}_t)\mathbf{G}_t(\mathbf{X}_t)^T$.  A similar approach can be used to sample from $p(\mathbf{x}_t | \mathbf{x_0})$ during training as follows


\begin{equation}
\mathbf{X}_t | \mathbf{X_0} = \mathbf{X}_0 + \frac{t}{T} \sqrt{\frac{1}{I_0}} \mathbf{A}^{+} D\{\frac{1}{\sqrt{\exp{(-\mathbf{A}\mathbf{X}_0)}}}\} \boldsymbol{\tilde{\epsilon}}_t\label{p_x_t_given_x0_2}
\end{equation}

\noindent where $\boldsymbol{\tilde{\epsilon}}_t$ is zero-mean identity-covariance white noise in the projection domain. The sampling algorithm, using the Euler-Maruyama method, is shown in Algorithm 2.

\begin{figure*}
    \centering
    \includegraphics[trim={40mm 35mm 40mm 30mm},clip,width=\textwidth]{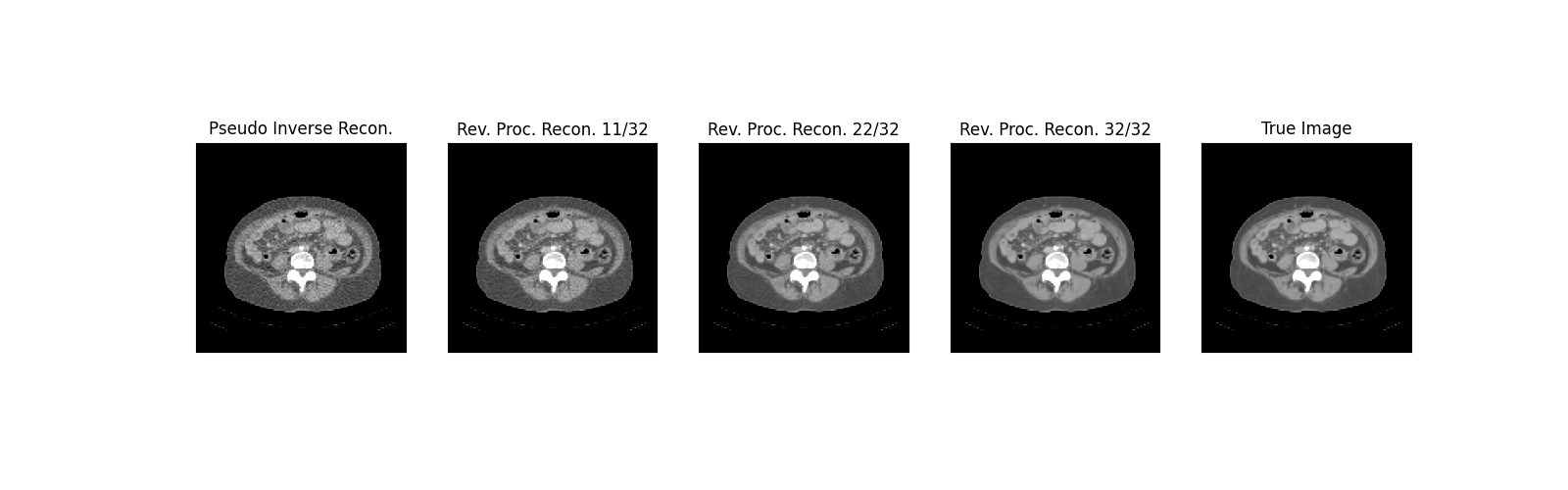}
    \caption{The pseudoinverse reconstructions are generated directly from the measurements and used to initialize the reverse diffusion process. The intermediate time points have the same noise correlations but lower noise magnitude than initial pseudoinverse CT reconstruction. }
    \label{fig:enter-label}
\end{figure*}

\begin{figure*}
    \centering
    \includegraphics[width=\textwidth]{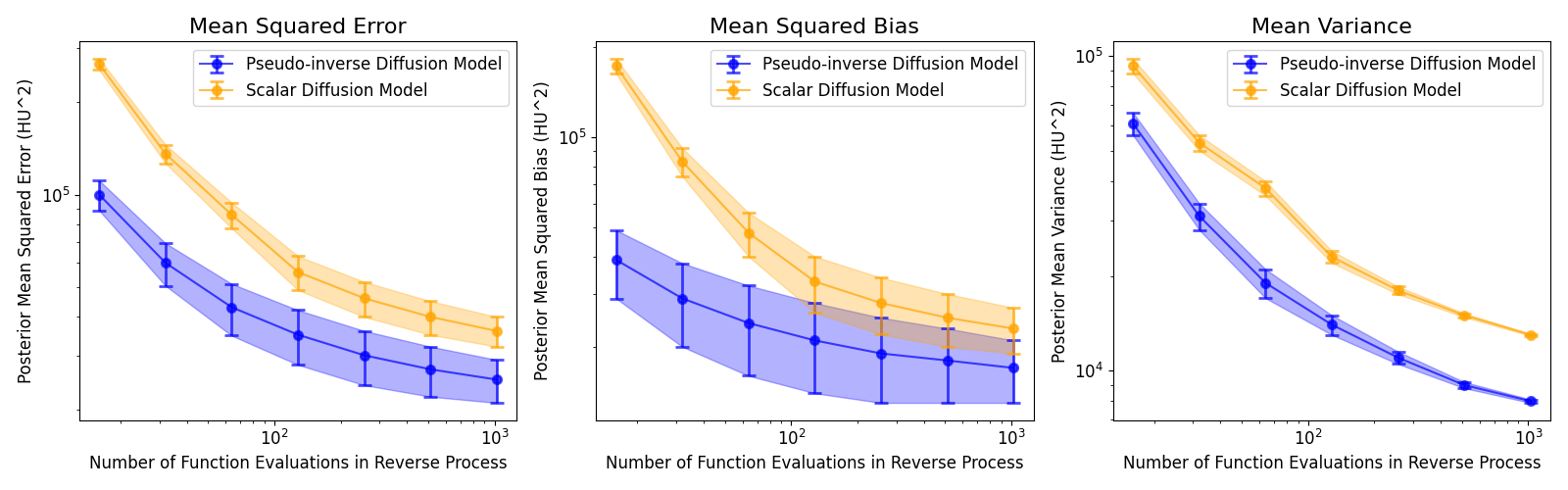}
    \caption{Performance metrics for posterior estimates of ground truth normal dose images given simulated low dose CT reconstructions. Orange curves show conditional diffusion models using a scalar diffusion process and blue curves show the proposed pseudoinverse diffusion models.  The solid line shows the mean and the error bars and filled region shows one standard deviation over all voxels and validation samples. }
    \label{fig:enter-label}
\end{figure*}

\subsection{Computational Experiment}
We conducted a comprehensive evaluation of our proposed method using a dataset from The Cancer Genome Atlas Liver Hepatocellular Carcinoma (TCGA-LIHC). This dataset consists of normal dose  abdominal CT images. We extracted 8000 training samples and 2000 validation samples of 256x256 axial cross sections 1.6mm voxel spacing which were treated as the ground truth. We simulated low dose CT measurements using the LEAP CT forward projector \cite{kim2023differentiable}. We set the number of view angles to 720 covering 360 degrees and modeled a single row of pixels with 1024 columns for the detector array. The pixel spacing was 2.0mm in both height and width. . The source-to-object distance was set at 600.0 mm, and the source-to-detector distance was 1200.0 mm. Projection domain Poisson was added corresponding to 10,000 photons per pixel per view. For the score matching neural network, we used an open source convolutional U-Net with 32 base channels, a depth of 4 down sampling layers, and channelwise self attention modules in the two deepest layers. We trained the model with a learning rate of 0.001 with the Adam optimizer for 1000 epochs and 128 batches per epoch and 16 images per batch \cite{kingma2014adam}. For a baseline, we trained a conditional diffusion model following the methods in \cite{batzolis2021conditional}. We trained the pseudoinverse diffusion models using Algorithm 1. 


\section{Results}
The results of our computational experiments are shown in Figures 3 and 4. In Figure 3, we showcased the reverse diffusion process initiated from pseudoinverse reconstructed images. The intermediate stages of this reverse process corresponded to varying radiation dose levels. Figure 4 presented a comprehensive quantitative analysis of the model's performance. The results show that pseudoinverse diffusion models can out-perform the conditional diffusion models, particularly for a low number of function evaluations. By breaking down mean squared error into mean squared bias and mean variance components, we can see that the majority of the improvement comes from lowering the posterior variance.


\section{Discussion}

The Moore-Penrose pseudoinverse reconstruction is not common due to stability concerns and computational limitations. The stability issue is related to a low condition number of $\mathbf{A}^T\mathbf{A}$. In the future, we are interested to explore penalized likelihood reconstruction using $(\mathbf{A}^T\mathbf{A} + \mathbf{R})^{-1} \mathbf{A}^T$, but this biased estimator is not compatible with our current model. The computational implementation of pseudoinverse reconstruction can also be a challenge. One method is to store $(\mathbf{A}^T\mathbf{A})^{-1}$ as a dense image domain matrix operator, as we did in our experiments, but this requires a large amount of memory. It is also possible to implement the pseudoinverse via iterative reconstruction methods, but it could require many iterations to converge. Filtered back projection (FBP) reconstruction is a promising approximation because it is computationally efficient in terms of both time and memory. Furthermore, FBP and similar analytical methods are based on the exact inverse in a spatially continuous framework \cite{Katsevich_2002}. In the future, we are interested to apply the FBP approximated approach in large three-dimensional images.

\section{Conclusion}

Our results demonstrate the efficacy of pseudoinverse diffusion models in reconstructing CT images, particularly from low-dose data. A key aspect of our method is the ability to maintain noise correlations and only reduce noise magnitude in the reverse process such that intermediate images correspond to the noise covariance of pseudoinverse reconstructions at various radiation dose levels. This characteristic addresses a common concern among radiologists regarding the altered texture in deep learning-based reconstructions. This advancement represents a significant stride in the realm of score based generative models for CT reconstruction.

\newpage

\bibliographystyle{IEEEtran}
\bibliography{report}
\end{document}